\documentclass[english]{article}
\usepackage[T1]{fontenc}
\usepackage[latin9]{inputenc}
\usepackage{xcolor}
\usepackage{units}
\usepackage{amsmath}
\PassOptionsToPackage{normalem}{ulem}
\usepackage{ulem}

\makeatletter

\providecolor{lyxadded}{rgb}{0,0,1}
\providecolor{lyxdeleted}{rgb}{1,0,0}

\DeclareRobustCommand{\lyxsout}[1]{\ifx\\#1\else\sout{#1}\fi}

\@ifundefined{date}{}{\date{}}
\usepackage[all]{xy}
\usepackage{placeins}
\usepackage{lineno}
\usepackage{flushend}
\usepackage{graphicx}
\usepackage{changepage}
\usepackage[titletoc,title]{appendix}
\usepackage{titlesec}
\usepackage{geometry}
\usepackage{cite}

\titleformat{\subsection}[runin]
  {\normalfont\bfseries}{[\thesubsection}{1pt}{]}

\makeatother

\usepackage{babel}
\begin{document}
\title{Assumption-free derivation of the Bell-type criteria of contextuality/nonlocality}
\author{Ehtibar N.\ Dzhafarov\\\\Purdue University, USA, ehtibar@purdue.edu}
\maketitle
\begin{abstract}
Bell-type criteria of contextuality/nonlocality can be derived without
any falsifiable assumptions, such as context-independent mapping (or
local causality), free choice, or no-fine-tuning. This is achieved
by deriving Bell-type criteria for inconsistently connected systems
(i.e., those with disturbance/signaling), based on the generalized
definition of contextuality in the Contextuality-by-Default approach,
and then specializing these criteria to consistently connected systems.

\end{abstract}

\section{Introduction}

The criteria (necessary and sufficient conditions) of contextuality/nonlocality
(usually, but not necessarily, in the form of inequalities) are at
the heart of the foundations of quantum physics. Since John Bell's
pioneering work, researchers have been interested in what assumptions
about nature one needs to justify these criteria. Several such assumptions
have been proposed. One of them is that measurements cannot be affected
by spacelike remote events (\emph{local causality}). Another assumption
is that experimenters choose what they measure independent of the
background events determining measurement outcomes (\emph{free choice}).
It has also been proposed that statistically identical measurement
outcomes should have identical ontological models (\emph{no-fine-tuning}).
This article shows that the criteria of contextuality/nonlocality can
be derived without any such assumptions.

Although our discussion is valid for essentially all possible systems
of random variables, we will use a system describing the Bohm's version of the Einstein-Podolsky-Rosen experiment
(EPR/B) \cite{Bohm1957,Bell1966,CHSH1969} as a throughout example. This allows us to avoid
technicalities needed in a more general exposition. The system in
question is

\begin{equation}
\begin{array}{|c|c|c|c||c|}
\hline R_{1}^{1} & R_{2}^{1} &  &  & c=1\\
\hline  & R_{2}^{2} & R_{3}^{2} &  & c=2\\
\hline  &  & R_{3}^{3} & R_{4}^{3} & c=3\\
\hline R_{1}^{4} &  &  & R_{4}^{4} & c=4\\
\hline\hline q=1 & q=2 & q=3 & q=4 & \textnormal{system }\mathcal{R}_{4}
\\\hline \end{array}\,.\label{eq:1}
\end{equation}

The random variables in (\ref{eq:1}) represent outcomes of spin measurements
of two entangled particles along direction $q$; in every trial Alice
chooses between directions $q=1$ and $3$, and Bob chooses between
$q=2$ and $4$ (indicated by subscripts of the random variables).
We call subscripts $q$ in $R_{q}^{c}$ the \emph{contents} of the corresponding
random variables. The four combinations of the Alice-Bob choices form
\emph{contexts} $c=1,\ldots,4$ (indicated by the superscripts of
the random variables). If the spin along an axis $q$ (i.e., content
$q$) is measured in context $c$, we write:
\begin{equation}
q\prec c.
\end{equation}

For instance $q=2\prec c=1$ but $q=3\not\prec c=1$. The particles
are spin-$\nicefrac{1}{2}$, so all the random variables in the system
are dichotomous, say $\pm1$.

The structure of this article is as follows. In Section \ref{fig2}, we do preliminary
work: we present a rigorous version of the traditional account of
the criteria for contextuality/nonlocality of systems of random variables
with no disturbance (we call them consistently connected systems),
and then we stipulate the traditional assumptions involved in the
formulation of these criteria. The main point of this paper is presented
in Section \ref{fig3}. There, we generalize the definition of contextuality/nonlocality
to inconsistently connected systems (those with disturbance or signaling),
and show that the generalized Bell-type criteria in this generalized
conceptual setting can be derived with no assumptions. Instead they
are derived from a classificatory definition of (non)contextuality
based on the notion of the difference between random variables. The traditional
Bell-type criteria then immediately follow by specializing the generalized
criteria to consistently connected systems, again, with no assumptions
about nature.

The term ``nonlocality'' has been criticized as misleading, unnecessary,
or even outright contradicting the tenets of quantum physics in several
recent publications \cite{Kupczynski2018,Khrennikov2020,Kupczynski2021}.
In this paper, however, the term is used to simply designate a special
case of contextuality, for systems where contexts are formed by spacelike
separated components. Barring experimental biases, such a system is
consistently connected, and its mathematical analysis does not differ
from that of other consistently connected systems \cite{AbramskyBrand2011,KDL2015}.
Singling nonlocal systems out, however, is justified due to their
special importance in quantum physics.

\section{Contextuality for Consistently Connected Systems}\label{fig2}

Consider a random variable $\Lambda$ and a measurable function $F$
mapping $\Lambda$ and $q$ into $\left\{ -1,1\right\} $. In contextuality/nonlocality
analysis of a system we are asking whether one can choose$\left(\Lambda,F\right)$
so that, for every context $c$,
\begin{equation}
\left(F\left(q,\Lambda\right):q\prec c\right)\overset{d}{=}\left(R_{q}^{c}:q\prec c\right),\label{eq:coupling property}
\end{equation}
where $\overset{d}{=}$ stands for ``has the same distribution as''.
In other words, the joint distribution of all $F\left(q,\Lambda\right)$
for a given context $c$ is the same as the joint distribution of
all $R_{q}^{c}$ in this context. Thus, for system $\mathcal{R}_{4}$
in (\ref{eq:1}),
\begin{equation}
\begin{array}{c}
\left(F\left(q=1,\Lambda\right),F\left(q=2,\Lambda\right)\right)\overset{d}{=}\left(R_{1}^{1},R_{2}^{1}\right),\\
\left(F\left(q=2,\Lambda\right),F\left(q=3,\Lambda\right)\right)\overset{d}{=}\left(R_{2}^{2},R_{3}^{2}\right),\\
etc.
\end{array}
\end{equation}

Note that the random variables $\left(R_{q}^{c}:q\prec c\right)$
in a given context $c$ are jointly distributed: e.g., the event $\left[R_{1}^{1}=x,R_{2}^{1}=y\right]$
is well-defined for any $x,y\in\left\{ -1,1\right\} $, and has a
probability assigned to it. At the same time, random variables in
different contexts, e.g., $R_{1}^{1}$ and $R_{1}^{2}$, or $R_{1}^{1}$
and $R_{2}^{2}$, do not have a joint distribution, as different contexts
are mutually exclusive \cite{DK2017}. However, all random variables
$F\left(q,\Lambda\right)$ are jointly distributed, because $\Lambda$
is one and the same for all $q$ and $c$. Therefore, the random variables
$S_{q}^{c}$ below form a (probabilisitic) \emph{coupling} of system
$\mathcal{R}_{4}$: 
\begin{equation}
\begin{array}{|c|c|c|c|}
\hline S_{1}^{1}=F\left(1,\Lambda\right) & S_{2}^{1}=F\left(2,\Lambda\right) &  & \\
\hline  & S_{2}^{2}=F\left(2,\Lambda\right) & S_{3}^{2}=F\left(3,\Lambda\right) & \\
\hline  &  & S_{3}^{3}=F\left(3,\Lambda\right) & S_{4}^{3}=F\left(4,\Lambda\right)\\
\hline S_{1}^{4}=F\left(1,\Lambda\right) &  &  & S_{4}^{4}=F\left(4,\Lambda\right)
\\\hline \end{array}\,.\label{eq:couplingg}
\end{equation}

Generally, a coupling 
 of a system of random variables $\mathcal{R}=\left\{ R_{q}^{c}:c\in C,q\in Q,q\prec c\right\} $
is a set of jointly distributed $S=\left\{ S_{q}^{c}:c\in C,q\in Q,q\prec c\right\} $
such that, for any $c\in C$, 
\begin{equation}
S^{c}=\left\{ S_{q}^{c}:q\in Q,q\prec c\right\} \overset{d}{=}\left\{ R_{q}^{c}:q\in Q,q\prec c\right\} =R^{c}.
\end{equation}
Note that any set of jointly distributed random variables is a random
variable. Therefore, $S$, $S^{c}$, and $R^{c}$ are random variables,
but $\mathcal{R}$ is not (hence the difference in notation).

In the coupling (\ref{eq:couplingg}), the octuple of the $S_{q}^{c}$-variables is jointly distributed,
and the within-context joint distributions of $S_{q}^{c}$ variables
are the same as the joint distributions of the corresponding $R_{q}^{c}$
variables. In addition, since $F\left(q,\Lambda\right)$ does not
depend on $c$, we have (with $\Pr$ standing for probability):
\begin{equation}
\Pr\left[S_{q}^{c}=S_{q}^{c'}\right]=1,
\end{equation}
for any $q\prec c,c'$ . In other words, the probabilities
\begin{equation}
\Pr\left[S_{1}^{1}=s_{1}^{1},S_{2}^{1}=s_{2}^{1},\ldots,S_{4}^{4}=s_{4}^{4},S_{1}^{4}=s_{1}^{4}\right]\label{eq:hidden vars}
\end{equation}
are well-defined for all $2^{8}$ octuples of $s_{q}^{c}\in\left\{ -1,1\right\} $.
These probabilities are subject to the following constraints:
\begin{enumerate}
\item For every fixed $\left(c,q,q'\right)$, such that $q,q'\prec c$, and
any fixed values $s_{q}^{c},s_{q'}^{c}\in\left\{ -1,1\right\} $,
\begin{equation}
\sum\iota_{\left[S_{q}^{c}=s_{q}^{c},S_{q'}^{c}=s_{q'}^{c}\right]}\Pr\left[S_{1}^{1}=s_{1}^{1},\ldots,S_{1}^{4}=s_{1}^{4}\right]=\Pr\left[R_{q}^{c}=s_{q}^{c},R_{q'}^{c}=s_{q'}^{c}\right],\label{eq::bunches}
\end{equation}
{where the coefficient $\iota_{\left[expression\right]}$ is the Boolean
indicator of whether the event} \mbox{$S_{1}^{1}=s_{1}^{1},\ldots,S_{1}^{4}=s_{1}^{4}$}
contains $expression$$;$
\item For any fixed $\left(c,c',q\right)$ such that $q\prec c,c'$, and
any fixed values $s_{q}^{c},s_{q}^{c'}\in\left\{ -1,1\right\} $,
\begin{adjustwidth}{-4.6cm}{0cm}
\begin{equation}
\sum\iota_{\left[S_{q}^{c}=s_{q}^{c},S_{q}^{c'}=s_{q}^{c'}\right]}\Pr\left[S_{1}^{1}=s_{1}^{1},\ldots,S_{1}^{4}=s_{1}^{4}\right]=\left\{ \begin{array}{lcc}
0 & \textnormal{if} & s_{q}^{c}\not=s_{q}^{c'}\\
\Pr\left[R_{q}^{c}=s_{q}^{c}\right]=\Pr\left[R_{q}^{c'}=s_{q}^{c}\right] & \textnormal{if} & s_{q}^{c}=s_{q}^{c'}
\end{array}\right..\label{eq:connections}
\end{equation}
\end{adjustwidth}

\end{enumerate}

This can be compactly presented as a linear programing (LP) task \cite{DCK2017,AbramskyBrand2011}:
\begin{equation}
\mathbf{MX}=\mathbf{P},\mathbf{X}\geq0,\label{eq:LP}
\end{equation}
where $\mathbf{X}$ is the vector of the $2^{8}$ probabilities (\ref{eq:hidden vars}),
$\mathbf{P}$ is the vector of the probabilities in the right-hand sides
of (\ref{eq::bunches}) and (\ref{eq:connections}), and $\mathbf{M}$
is a Boolean incidence matrix, whose entries are the $\iota$-coefficients
in  (\ref{eq::bunches}) and (\ref{eq:connections}). The condition
$\mathbf{X}\geq0$ (componentwise) ensures that the solution $\mathbf{X}$,
if it exists, consists of numbers interpretable as probabilities (the
summation of these values to 1 in  (\ref{eq:LP}) is ensured).

Denoting by $\mathsf{LP}\left[\mathbf{M},\mathbf{P}\right]$ a Boolean
function that equals 1 if and only if a solution $\mathbf{X}$ exists,
it is well-known that $\mathsf{LP}\left[\mathbf{M},\mathbf{P}\right]$
is computable in polynomial time \cite{Karmarkar1984}. Consequently,
\begin{equation}
\mathsf{LP}\left[\mathbf{M},\mathbf{P}\right]=1\label{eq:LPcriterion}
\end{equation}
can be taken as a criterion of noncontextuality/locality. As an optional
step, by a facet enumeration algorithm, this criterion can be presented
in the form of inequalities and equations involving moments of the
distributions within contexts. For system $\mathcal{R}_{4}$, this
yields \cite{CHSH1969,Fine1982,DK2013}: 
\begin{adjustwidth}{-4.6cm}{0cm}
\begin{equation}
\left|\left\langle R_{1}^{1}R_{2}^{1}\right\rangle +\left\langle R_{2}^{2}R_{3}^{2}\right\rangle +\left\langle R_{3}^{3}R_{4}^{3}\right\rangle +\left\langle R_{4}^{4}R_{1}^{4}\right\rangle -2\min\left(\left\langle R_{1}^{1}R_{2}^{1}\right\rangle ,\left\langle R_{2}^{2}R_{3}^{2}\right\rangle ,\left\langle R_{3}^{3}R_{4}^{3}\right\rangle ,\left\langle R_{4}^{4}R_{1}^{4}\right\rangle \right)\right|\leq2,\label{eq:CHSH}
\end{equation}
\end{adjustwidth}

\begin{equation}
\left\langle R_{1}^{1}\right\rangle =\left\langle R_{1}^{4}\right\rangle ,\left\langle R_{2}^{1}\right\rangle =\left\langle R_{2}^{2}\right\rangle ,\left\langle R_{3}^{2}\right\rangle =\left\langle R_{3}^{3}\right\rangle ,\left\langle R_{4}^{3}\right\rangle =\left\langle R_{4}^{4}\right\rangle .\label{eq:nondisturbance}
\end{equation}

Equalities (\ref{eq:nondisturbance}) present the condition of consistent
connectedness.

What ontological assumptions have we made in the foregoing? Let us
note first that we need no assumptions to derive the criterion $\mathsf{LP}\left[\mathbf{M},\mathbf{P}\right]=1$
or its equivalents from~\mbox{(\ref{eq:coupling property}).} The derivation
of the criterion is by straightforward linear programming, optionally
complemented by facet enumeration. However, we need certain assumptions
about nature to justify the plausibility of using the function $F\left(q,\Lambda\right)$,
which is the starting point of the derivation. Namely, we must have
assumed that the mapping does not contain context $c$ among its arguments.
For the EPR/Bohm experiment, following Bell \cite{BellLaNouvelle},
this assumption can be called \emph{local causality}, because dependence
of the mapping on $c$ can be interpreted as dependence of measurement
outcomes on spacelike-remote settings. More generally, however, this
can be called \emph{context-independent mapping}, in order to also
include the Kochen-Specker type contextuality \cite{KS1967}, when
measurements in the same context are not spacelike separated. We must
have also assumed that $\Lambda$ in $F\left(q,\Lambda\right)$ is
one and the same for all $q$ and for all $c$. This is called the
assumption of \emph{free choice}, or \emph{statistical independence
}(of measurements and settings). In relation to system $\mathcal{R}_{4}$,
the necessity of adding the free choice assumption to the local causality
assumption was pointed out to John Bell by Shimony, Horne, and Clauser
in their 1985 interchange \cite{BellvsSHC1985,Norsen2011}. The relationship
between free choice and local causality (more generally, context-independent
mapping) is an interesting issue, but it is discussed elsewhere \cite{DzhafarovNew}.

{One can replace both the assumption of context-independent mapping
and the free choice assumption with the assumption proposed by Cavalcanti,
called \emph{no-fine-tuning}~\mbox{\cite{Cavalcanti2018,PearlCavalcanti}}.}
For our purposes it can be formulated thus: if two random variables
sharing a content in different contexts have the same distribution, their representations in the form
\begin{equation}
F\left(\textnormal{some parameters},\text{\textnormal{some random variables}}\right)
\end{equation}
should be identical. This is an attractive alternative to context-independent
mapping, because the latter is not especially compelling in the cases
of Kochen-Specker-type contextuality, when contexts are not defined
by spacelike remote settings. Note that the no-fine-tuning can also
be considered a principle of theory construction, essentially a conceptual
parsimony principle, rather than an ontological assumption.

As it turns out, however, by generalizing the notion of (non)contextuality
to include inconsistently connected systems, one can avoid the necessity
of making any of these, or other, falsifiable assumptions. The no-fine-tuning
assumption (or parsimony principle) is a consequence of specializing
this general definition to consistently connected systems.

\section{Contextuality in Inconsistently Connected Systems}\label{fig3}

The generalization follows the broadening of the class of systems
of random variables amenable to contextuality analysis. As one can
see in (\ref{eq:nondisturbance}), the criterion $\mathsf{LP}\left[\mathbf{M},\mathbf{P}\right]=1$
can be satisfied only if the system of random variables is consistently
connected: random variables measuring the same property in different
contexts have the same distribution. We will now drop this constraint,
and allow for inconsistent connectedness. In particular, we allow
for signaling between Alice and Bob if their measurements are timelike
separated.

We begin with the maximally lax representation
\begin{equation}
R_{q}^{c}=\gamma\left(q,c,\lambda_{q}^{c}\right),\label{eq:initgen}
\end{equation}
in which both the outcome of the measurement and the background random
variable are allowed to depend on both $q$ and $c$. This representation
obviously holds for any system of random variables. Now, because all
$R_{q}^{c}$ in a given context $c$ are jointly distributed, it follows
that all $\lambda_{q}^{c}$ in this context are jointly distributed.
Therefore, there is a random variable $\lambda^{c}$ of which $\lambda_{q}^{c}$
for all $q\prec c$ are functions. As a result, we get a seemingly
more restrictive but still universally applicable representation:

\begin{equation}
R_{q}^{c}=g\left(q,c,\lambda^{c}\right).\label{eq:secondgen}
\end{equation}

The remaining dependence of $\lambda^{c}$ on context $c$ can also 
be eliminated, by the following reasoning. Let us form an arbitrary
coupling
\begin{equation}
\Lambda\overset{d}{=}\left(\lambda^{c}:\textnormal{all }c\right),
\end{equation}
e.g., couple all $\lambda^{c}$ independently. Then:
\begin{equation}
R_{q}^{c}\overset{d}{=}g\left(q,c,\mathrm{Proj}_{c}\left(\Lambda\right)\right)\overset{}{=}G\left(q,c,\Lambda\right),\label{eq:final universal}
\end{equation}
where $\mathrm{Proj}_{c}$ is the $c$th projection function. This
representation (with $\Lambda$ one and the same for all $q$ and
$c$) is also universally applicable. Note that in  (\ref{eq:final universal}) we
only have $R_{q}^{c}\overset{d}{=}G\left(q,c,\Lambda\right)$, rather
than $R_{q}^{c}=G\left(q,c,\Lambda\right)$, because the latter would
make $R_{q}^{c}$ jointly distributed across mutually exclusive contexts
$c$ (which is nonsensical).

It remains to define (non)contextuality in this generalized conceptual
setting. In the contextuality-by-default approach (CbD) \cite{DCK2017,DK2017,Dzh2019,KDL2015,DKC2020},
the system is considered noncontextual if $\left(\Lambda,G\right)$
in (\ref{eq:final universal}) can be chosen so that the probability:
\begin{equation}
\Pr\left[G\left(q,c,\Lambda\right)=G\left(q,c',\Lambda\right)\right]\label{eq:pair}
\end{equation}
is the maximal possible, for all $\left(q,c,c'\right)$ such that $q\prec c,c'$.
For dichotomous random variables, this means
\begin{equation}
\Pr\left[G\left(q,c,\Lambda\right)=G\left(q,c',\Lambda\right)=1\right]=\min\left\{ \begin{array}{c}
\Pr\left[R_{q}^{c}=1\right]\\
\\
\Pr\left[R_{q}^{c'}=1\right]
\end{array}\right..
\end{equation}

The rationale for this definition is as follows. The maximal probability
with which $G\left(q,c,\Lambda\right)$ and $G\left(q,c',\Lambda\right)$
can be made to coincide shows how similar the random variables $R_{q}^{c}$
and $R_{q}^{c'}$ are if taken as an isolated pair, ``out of their
contexts.'' Denoting by $p_{q}^{c}$ the value of $\Pr\left[F\left(q,c,\Lambda\right)=1\right]=\Pr\left[R_{q}^{c}=1\right]$,
the maximum probability of $F\left(q,c,\Lambda\right)=F\left(q,c',\Lambda\right)$
is $1-\left|p_{q}^{c}-p_{q}^{c'}\right|$, and it is achieved if and
only if the joint distribution of $S_{q}^{c}=F\left(q,c,\Lambda\right)$
and $S_{q}^{c'}=F\left(q,c',\Lambda\right)$ is (assuming
$p_{q}^{c}\leq p_{q}^{c'}$):
\begin{equation}
\begin{array}{|c||c|c||c|}
\hline  & S_{q}^{c}=1 & S_{q}^{c}=-1 & \\
\hline\hline S_{q}^{c'}=1 & p_{q}^{c} & 0 & p_{q}^{c}\\
\hline S_{q}^{c'}=-1 & p_{q}^{c'}-p_{q}^{c} & 1-p_{q}^{c'} & 1-p_{q}^{c}\\
\hline\hline  & p_{q}^{c'} & 1-p_{q}^{c'} & 
\\\hline \end{array}\,.\label{eq:maximal}
\end{equation}

The existence of a coupling of the system in which all pairs $\left(S_{q}^{c},S_{q}^{c'}\right)$
have this joint distribution indicates that the way $R_{q}^{c}$ and $R_{q}^{c'}$ are related to other random variables in the
corresponding contexts leaves their dissimilarity intact. Otherwise,
the contexts of the system ``force'' some of the pairs of $R_{q}^{c}$
and $R_{q}^{c'}$ to be more dissimilar than they are when taken in
isolation.

The Bell-type criterion for this generalized definition is also determined
by the LP problem  (\ref{eq:LP}). The only difference is that the
part of vector $\mathbf{P}$ determined by  (\ref{eq:connections})
is replaced with the probabilities given in  (\ref{eq:maximal}):
\begin{equation}
\sum\iota_{\left[S_{q}^{c}=s_{q}^{c},S_{q}^{c'}=s_{q}^{c'}\right]}\Pr\left[S_{1}^{1}=s_{1}^{1},\ldots,S_{1}^{4}=s_{1}^{4}\right]=\left\{ \begin{array}{lcc}
p_{q}^{c} & \textnormal{if} & s_{q}^{c}=s_{q}^{c'}=1\\
1-p_{q}^{c'} & \textnormal{if} & s_{q}^{c}=s_{q}^{c'}=-1\\
 & etc.
\end{array}\right..
\end{equation}

Note that this definition of (non)contextuality is purely classificatory,
it does not involve any assumptions about nature.

Suppose now that the system is consistently connected. Then the maximal
probability in  (\ref{eq:pair}) is 1, i.e.,  (\ref{eq:maximal}) holds
with $p_{q}^{c}=p_{q}^{c'}$, and the system is noncontextual if and
only if $\left(\Lambda,G\right)$ can be chosen so that:
\begin{equation}
\Pr\left[G\left(q,c,\Lambda\right)=G\left(q,c',\Lambda\right)\right]=1,\label{eq:specialization}
\end{equation}
whenever $q\prec c,c'$. But the latter means that
\begin{equation}
G\left(q,c,\Lambda\right)\equiv F\left(q,\Lambda\right).\label{eq:final}
\end{equation}

We have thus arrived at the same representation as in the previous
section, but without assuming context-independent mapping, free choice,
or no-fine-tuning.

\section{Conclusions}

We have seen that if one does not make any constraining assumptions
about how a measurement outcome in a system depends on settings, local
or remote, and if one adopts the CbD definition of generalized (non)contextuality,
the derivation of the traditional criteria of (non)contextuality follows
with no ontological assumptions, by simply specializing the definition
in question to consistently connected systems.

Let us examine possible doubts about the validity of our analysis.

(1) Is not 
 the property of consistent connectedness from which we derive \mbox{ (\ref{eq:specialization})} and \mbox{(\ref{eq:final})} an assumption? If
the distributions of the random variables are only known to us on
a sample level, then it is indeed an assumption, as any other statistical
hypothesis. However, the discussion in this article deals with the
systems known to us precisely, as random variables rather than samples.
For example, a standard quantum mechanical computation can yield the
precise joint probabilities for the pairs of variables in each context
of system $\mathcal{R}_{4}$.

(2) By 
 writing $R_{q}^{c}=\gamma\left(q,c,\lambda_{q}^{c}\right)$,
have we not made the assumption of ``outcome determinism''? The
latter means that the values of the relevant arguments uniquely determine
the value of $R_{q}^{c}$. What if $q,c$, and $\lambda_{q}^{c}$
only determine the distribution of the random variable $R_{q}^{c}$
rather than its value? This possibility, however, amounts to introducing
yet another random variable among the arguments of the function determining
the value of $R_{q}^{c}$:
\begin{equation}
R_{q}^{c}=\gamma\left(q,c,\lambda_{q}^{c},\mu_{q}^{c}\right).
\end{equation}
This obviously reduces to the initial representation on renaming $\left(\lambda_{q}^{c},\mu_{q}^{c}\right)$
into $\lambda_{q}^{c}$. This observation is, in fact, a short (and
generalized) version of a theorem established as early as 1982
by Arthur Fine \cite{Fine1982}.

(3) Is 
 not the generalized CbD definition of contextuality a form of
the no-fine-tuning assumption? This is clearly not true for the original
no-fine-tuning assumption \cite{Cavalcanti2018,PearlCavalcanti},
as the latter is confined to consistently connected systems. However,
it has been shown by M. Jones \cite{Jones2019} that it is possible
to generalize the no-fine-tuning assumption to become a principle
that forbids ``hidden influences'' in ontological models of a system.
``Hidden influences'' mean dependence of measurement outcomes on
factors that do not influence the distributions of these outcomes.
Maximizing the probability of $G\left(q,c,\Lambda\right)=G\left(q,c',\Lambda\right)$
ensures that the entire difference between the influences of $c$
and $c'$ on the measurement of $q$ is reflected in the difference
of their distributions. However, in CbD, this is a consequence of defining
(non)contextuality of a system in terms of differences between content-sharing
random variables, rather than an assumption about nature or a principle
for constructing plausible ontological theories.

(4) Finally, let us address the question about the
distinction we make between ontological assumptions and a classificatory
definition. Is it defensible? Is not any assumption convertible into a definition and
vice versa? Specifically, the traditional analysis of contextuality
can be presented as a definition according to which a (consistently
connected) system is noncontextual if a representation
\begin{equation}
\left(R_{q}^{c}:q\prec c\right)\overset{d}{=}\left(F\left(q,\Lambda\right):q\prec c\right)\label{eq:last}
\end{equation}
exists, and it is noncontextual otherwise. Not denying this obvious possibility, it is nevertheless reasonable
to ask how such a definition is motivated. The assumptions of context-independent
mapping, free choice, and no-fine-tuning serve to provide this motivation. (Let us note in passing that most of the traditional analyses of contextuality confuse the distributional
equation \mbox{ (\ref{eq:last})} with the equality 
 $R_{q}^{c}=F\left(q,\Lambda\right)$. The ensuing logical problems are discussed in \cite{DK2017,Dzh2019}.)
In our analysis, we begin with a hidden variable model that cannot
be empirically false, $R_{q}^{c}=\gamma\left(q,c,\lambda_{q}^{c}\right)$,
and reduce it to $R_{q}^{c}\overset{d}{=}G\left(q,c,\Lambda\right)$,
that cannot be false either. Then we pose the question about the difference
between random variables $R_{q}^{c}$ and $R_{q}^{c'}$, measuring
the same content in different contexts. Asked about the motivation
for this question, the obvious answer is that we are interested in
the dependence of measurements on their contexts. The maximal probability
of $S_{q}^{c}=S_{q}^{c'}$ in the coupling
\begin{equation}
\left(S_{q}^{c}=F\left(q,c,\Lambda\right),S_{q}^{c'}=F\left(q,c',\Lambda\right)\right)
\end{equation}
provides the answer for $R_{q}^{c}$ and $R_{q}^{c'}$ taken in isolation
from their contexts. Finally, we ask the question if thus measured
differences between $R_{q}^{c}$ and $R_{q}^{c'}$ are compatible
with their respective contexts, by finding out if the maximal probability
of $S_{q}^{c}=S_{q}^{c'}$ remains the same or decreases within the
overall couplings of the system. If asked about the motivation for
this question, the answer is that the situations when these maximal
probabilities decrease (for some $R_{q}^{c}$ and $R_{q}^{c'}$) indicate
a special form of dependence of measurements on their contexts, beyond
the difference of their distributions. There seems to be no assumptions
about nature that would pertain to this distinction (see point 3 above).
To summarize, the point we make in this paper is not based on the
admittedly blurry distinction between definitions and assumptions.
Rather, it is based on dealing with a representation, $\gamma\left(q,c,\lambda_{q}^{c}\right)$
or $G\left(q,c,\Lambda\right)$, that is universally applicable, and
does not therefore involve any ontological assumptions.

\paragraph*{Acknowledgments}

The author is grateful to Mattew Jones for critically discussing the
earlier version of this paper.

\paragraph*{Conflict of interest}

The author declares no conflict of interest.

\end{document}